\pdfoutput=1

\documentclass[11pt]{article}

\usepackage{ACL2023}

\usepackage{times}
\usepackage{latexsym}

\usepackage[T1]{fontenc}

\usepackage[utf8]{inputenc}

\usepackage{microtype}

\usepackage{inconsolata}

\usepackage{graphicx}
\usepackage{amsmath}
\usepackage{multirow}
\usepackage{float}
\usepackage{microtype}
\usepackage{booktabs}
\usepackage{xcolor}
\usepackage[table]{xcolor} 
\newcommand{\redact}[1]{\textcolor{black}}
\usepackage{adjustbox} 
\usepackage{enumitem}
\usepackage{lscape}

\title{\textsc{CodeEval: A pedagogical approach for targeted evaluation of code-trained Large Language Models}}

\author{
Danny Brahman, Mohammad H. Mahoor\\ 
Department of Computer Science \\
University of Denver\\ Denver, USA \\
\texttt{danny.brahman@du.edu}\quad
\texttt{mohammad.mahoor@du.edu}
}
\begin{document}
\maketitle
\begin{abstract}
Large Language Models (LLMs) are predominantly assessed based on their common sense reasoning, language comprehension, and logical reasoning abilities. While models trained in specialized domains like mathematics or coding have demonstrated remarkable advancements in logical reasoning, there remains a significant gap in evaluating their code generation capabilities. Existing benchmark datasets fall short in pinpointing specific strengths and weaknesses, impeding targeted enhancements in models' reasoning abilities to synthesize code. To bridge this gap, our paper introduces an innovative, pedagogical benchmarking method that mirrors the evaluation processes encountered in academic programming courses. We introduce CodeEval, a multi-dimensional benchmark dataset designed to rigorously evaluate LLMs across 24 distinct aspects of Python programming. The dataset covers three proficiency levels—beginner, intermediate, and advanced—and includes both class-based and function-based problem types with detailed problem specifications and comprehensive test suites. To facilitate widespread adoption, we also developed RunCodeEval, an open-source execution framework that provides researchers with a ready-to-use evaluation pipeline for CodeEval. RunCodeEval handles test execution, context setup, and metrics generation, enabling researchers to quickly obtain detailed insights into model strengths and weaknesses across complexity levels, problem types, and programming categories. This combination enables targeted evaluation and guides improvements in LLMs' programming proficiencies.
\end{abstract}

\section{Introduction}
\begin{table*}
\centering
\begin{adjustbox}{max width=\textwidth}
\begin{tabular}{m{0.4\textwidth} m{0.3\textwidth} m{0.05\textwidth} m{0.2\textwidth} m{0.05\textwidth} m{0.1\textwidth} m{0.1\textwidth} m{0.1\textwidth} m{0.3\textwidth}}
\toprule
\textbf{Dataset} & \textbf{Focus} & \textbf{\# Tasks} & \textbf{Languages} & \textbf{Has test cases} & \textbf{Granularity} & \textbf{Hand-curated} & \textbf{Has testing framework} & \textbf{Input} \\ \midrule
CodeSearchNet \cite{husain2019codesearchnet} &  Various downstream tasks & ~2M & Go, Java, JavaScript, PHP, Python, Ruby & No & Function & No & No & - \\ \hline
GCJ \cite{ullah2019cyber} & Competitional & 2.4M & C++, C, Python, Java  & No & None & No & No  & - \\ \hline
CodeNet \cite{puri2021codenet} & Various downstream tasks & 13M & C++, C, Python, Java, Ruby, C\#  & No & None & No & No & - \\ \hline
CodeContests \cite{li2022competition} & Competitional & 13M & C++, Python, Java  & Yes & None & No & No & NL + input-output pairs \\ \hline
HumanEval \cite{chen2021evaluating} & Untargeted evaluation & 164 & Python  & Yes & Function & Yes & No & NL + Function signature + input-output pairs \\ \hline
MBPP \cite{austin2021program} & Entry level programming & 974 & Python  & Yes & Function & Yes & No & NL  \\ \hline
MathQA-Python \cite{amini2019mathqa} & Statement-level evaluation & 2,985 & Python  & No & Statement & Yes & No  & NL \\ \hline
APPS \cite{hendrycks2021measuring} & Competitional & 232,421 & Python  & Yes & None & No & No & NL + input-output pairs \\ \hline
CoderEval \cite{yu2024codereval} & Competitional & 230 & Python, Java  & No & Function & No & No & NL + Function signature \\ \hline
ClassEval \cite{du2023classeval} & Untargeted evaluation & 100 & Python & Yes & Class & Yes & No  & Class skeleton\\ \hline
\rowcolor{lightgray}
CodeEval & Targeted evaluation & 602 & Python  & Yes & Class + Function & Yes & Yes  & NL\\ \hline
\end{tabular}
\end{adjustbox}
\caption{Comparison of Existing Datasets with CodeEval}
\label{tab:benchmark-datasets}
\end{table*}
\begin{table}[h]
    \centering
    \scalebox{0.80}{
    \begin{tabular}{lc}
\toprule
\textbf{Characteristic} & \textbf{Value} \\
\midrule
\multicolumn{2}{c}{Basic Metrics} \\
\midrule
Total Problems & 602 \\
Programming Categories & 24 \\
Total Test Cases & 1,471 \\
Average Problems per Category & 25.1 \\
Average Tests per Problem & 2.4 \\
Average Test Coverage & 99.1\% \\
\midrule
\multicolumn{2}{c}{Problem Type Distribution} \\
\midrule
Function Problems & 374 (62.1\%) \\
Class Problems & 228 (37.9\%) \\
\midrule
\multicolumn{2}{c}{Complexity Distribution} \\
\midrule
Beginner & 135 (22.4\%) \\
Advanced & 264 (43.9\%) \\
Intermediate & 203 (33.7\%) \\
\midrule
  \multicolumn{2}{c}{Category Distribution} \\
  \midrule
  \multicolumn{2}{c}{24 categories, 20-48 problems each} \\
  \multicolumn{2}{c}{(see Table \ref{tab:category_distribution} in Appendix \ref{app:dataset_details})} \\
    \bottomrule
    \end{tabular}}
    \caption{Statistics of the Codeeval dataset}
    \label{tab:dataset_overview}
\end{table}
\begin{table*}[ht]
\centering
\resizebox{\textwidth}{!}{
\begin{tabular}{l|c|ccc|cc}
\toprule
\textbf{Model} & \textbf{Overall} & \multicolumn{3}{c|}{\textbf{Complexity}} & \multicolumn{2}{c}{\textbf{Problem Type}} \\
 & Score & Beginner & Intermediate & Advanced & Function & Class \\
\midrule
gpt-4.1-2025-04-14 & $90.5 \pm 2.1$ & $93.8 \pm 3.5$ & $94.7 \pm 2.4$ & $82.8 \pm 4.8$ & $91.4 \pm 2.5$ & $89.0 \pm 3.8$ \\
o3-mini-2025-01-31 & $89.5 \pm 2.3$ & $92.1 \pm 3.8$ & $93.3 \pm 2.8$ & $82.7 \pm 4.9$ & $90.9 \pm 2.6$ & $87.1 \pm 4.2$ \\
Qwen/Qwen3-235B-A22B-Instruct-2507 & $88.9 \pm 2.3$ & $93.8 \pm 3.2$ & $93.2 \pm 2.7$ & $80.1 \pm 5.1$ & $90.1 \pm 2.7$ & $87.1 \pm 4.1$ \\
o4-mini-2025-04-16 & $88.0 \pm 2.4$ & $94.0 \pm 3.4$ & $92.7 \pm 2.9$ & $78.1 \pm 5.4$ & $91.6 \pm 2.5$ & $82.2 \pm 4.7$ \\
claude-sonnet-4-20250514 & $86.8 \pm 2.5$ & $94.1 \pm 3.6$ & $91.5 \pm 3.1$ & $76.0 \pm 5.6$ & $88.6 \pm 3.0$ & $83.9 \pm 4.6$ \\
meta-llama/Llama-4-Maverick-17B-128E-Instruct & $86.5 \pm 2.5$ & $92.5 \pm 4.1$ & $87.6 \pm 3.8$ & $81.3 \pm 4.9$ & $88.9 \pm 2.9$ & $82.7 \pm 4.7$ \\
command-a-03-2025 & $85.9 \pm 2.5$ & $91.3 \pm 4.2$ & $89.0 \pm 3.4$ & $78.4 \pm 5.2$ & $87.1 \pm 3.0$ & $84.0 \pm 4.5$ \\
claude-opus-4-20250514 & $84.7 \pm 2.7$ & $93.8 \pm 3.5$ & $88.1 \pm 3.7$ & $74.2 \pm 5.6$ & $88.6 \pm 2.9$ & $78.4 \pm 5.1$ \\
grok-2-vision-1212 & $83.3 \pm 2.8$ & $93.1 \pm 3.7$ & $85.5 \pm 4.0$ & $73.8 \pm 5.8$ & $87.8 \pm 3.0$ & $75.8 \pm 5.3$ \\
meta-llama/Llama-4-Scout-17B-16E-Instruct & $82.8 \pm 2.8$ & $94.4 \pm 3.2$ & $84.0 \pm 4.2$ & $73.5 \pm 5.6$ & $86.7 \pm 3.1$ & $76.5 \pm 5.3$ \\
gemini-2.0-flash & $80.8 \pm 3.0$ & $92.1 \pm 3.9$ & $83.6 \pm 4.3$ & $69.6 \pm 6.0$ & $83.4 \pm 3.5$ & $76.5 \pm 5.4$ \\
gpt-3.5-turbo & $80.0 \pm 3.0$ & $89.1 \pm 4.7$ & $83.1 \pm 4.2$ & $69.9 \pm 6.0$ & $82.7 \pm 3.5$ & $75.6 \pm 5.4$ \\
Qwen/Qwen3-4B & $74.0 \pm 3.3$ & $89.9 \pm 4.5$ & $77.5 \pm 4.7$ & $58.9 \pm 6.4$ & $79.3 \pm 3.7$ & $65.2 \pm 6.0$ \\
command-r-08-2024 & $54.5 \pm 3.9$ & $58.8 \pm 8.1$ & $61.0 \pm 5.7$ & $43.0 \pm 6.6$ & $47.5 \pm 4.9$ & $65.9 \pm 5.9$ \\
command-r-plus-08-2024 & $21.1 \pm 3.2$ & $12.6 \pm 5.6$ & $26.0 \pm 5.2$ & $20.3 \pm 5.5$ & $3.3 \pm 1.8$ & $50.2 \pm 6.4$ \\
\bottomrule
\end{tabular}}
\caption{Comprehensive Model Performance on CodeEval Benchmark. Overall accuracy scores (\%) with 95\% confidence intervals across 15 state-of-the-art LLMs, broken down by complexity levels (beginner, intermediate, advanced) and problem types (Function vs Class-based). See Appendix \ref{app:model_info} for model information.}
\label{tab:comprehensive_performance}
\end{table*}

Large Language Models (LLMs) trained on code have shown remarkable logical reasoning abilities, yet current evaluation methods remain limited. Existing benchmarks assess functional correctness \cite{chen2021evaluating, hendrycks2021measuring} but fail to provide detailed insights into models' specific strengths and weaknesses, hindering targeted improvements. Moreover, these benchmarks focus on individual aspects (complexity, function-based, or class-based problems) rather than multi-dimensional evaluation.

We address these gaps with CodeEval, a pedagogical benchmark dataset of 602 hand-crafted Python problems spanning 24 programming categories across three complexity levels. Our approach uniquely combines function and class-based problems with context-aware test cases that enable complex evaluation scenarios. To make this benchmark practically usable, we developed RunCodeEval, an execution framework that leverages CodeEval's problems and test cases to generate comprehensive evaluation results. RunCodeEval handles the complex task of executing LLM solutions against CodeEval's test suite, providing granular analysis across problem categories, complexity levels, and problem types, with partial credit scoring and detailed error analysis for targeted model improvement.

Our contributions are as follows:
\begin{enumerate}[noitemsep, topsep=0pt]
\item Multi-dimensional Evaluation: While existing benchmarks focus on individual aspects such as complexity, function-based problems, or class-based problems, none provide a unified evaluation across all these dimensions. We introduce the first benchmark that provides multi-dimensional assessment of model performance across three levels of complexity and two distinct problem types (functional and class-based), covering a broad spectrum of Python programming concepts.

\item Hand-curated Benchmark: We developed a novel hand-curated benchmark dataset, CodeEval, specifically designed to enable targeted  analysis of model performance in synthesizing Python code.

\item Open-sourcing CodeEval and RunCodeEval: In addition to the CodeEval dataset, we have created and open-sourced RunCodeEval, an execution framework that operationalizes the CodeEval benchmark which is a non-trivial task given the complex nature of context-aware test cases in CodeEval. RunCodeEval solves this usability challenge by providing a complete evaluation pipeline that executes LLM solutions against CodeEval's comprehensive test suite, incorporating partial credit scoring and context-aware test execution logic. (Refer to Appendix \ref{app:access} for accessing CodeEval and RunCodeEval)

\item \textbf{Complete Reproducibility:} Complete code for generating LLM solutions across all 15 evaluated models will be made openly available to ensure full experimental reproducibility.
\end{enumerate}

\section{Related Datasets}
\label{related_work}
Early code evaluation datasets include CodeXGLUE \cite{lu2021codexglue}, a comprehensive collection covering tasks like clone detection, code completion, and text-to-code generation. Large-scale datasets such as GCJ \cite{ullah2019cyber} (2.4M samples, 332 problems), CodeNet \cite{puri2021codenet} (13M samples, 4,053 problems), and CodeContests \cite{li2022competition} (13,328 competition problems with test cases) provide extensive code repositories across multiple programming languages, primarily C++, Python, and Java.

Code generation benchmarks include HumanEval \cite{chen2021evaluating} (164 hand-curated problems with test cases), MBPP \cite{austin2021program} (974 entry-level Python functions), and APPS \cite{hendrycks2021measuring} (10,000 problems across three difficulty levels). Recent specialized datasets include Codereval \cite{yu2024codereval} (230 non-self-contained functions reflecting real-world dependencies), ClassEval \cite{du2023classeval} (class-based tasks with unique skeleton input style), mHumanEval \cite{raihan2025mhumaneval} (multilingual extension of HumanEval with prompts in 204 natural languages), and LiveCodeBench \cite{jain2024livecodebench} (continuously updated benchmark from coding contests to prevent contamination). However, these benchmarks focus on individual evaluation aspects and lack comprehensive multi-dimensional analysis across problem types, complexity levels, and programming concepts simultaneously.

To address these gaps, we developed CodeEval, a novel benchmarking dataset that surpasses existing benchmarks in several key ways:
\begin{enumerate}[noitemsep, topsep=1pt]
    \item \textbf{Hand-Curated Quality}: CodeEval consists of 602 carefully hand-crafted problems to ensure high quality and pedagogical value. Unlike datasets automatically scraped from online sources, each problem in CodeEval is deliberately designed to test specific programming concepts with comprehensive context-aware test coverage, clear problem statements, and canonical solutions.
    \item \textbf{Multi-dimensional Evaluation}: CodeEval includes both function-based and class-based problems across three complexity levels, offering a balanced evaluation of models' ability to synthesize diverse code structures. Unlike existing benchmarks, CodeEval is the first to provide multi-dimensional coverage across all three dimensions of evaluation.
	\item \textbf{Conceptual Coverage}: CodeEval explicitly targets specific Python concepts, such as data structures, recursion, and asynchronous programming, offering clarity on what is being tested. This level of detail is often missing in other datasets.
	\item \textbf{Targeted Evaluation}: CodeEval enables granular performance analysis by breaking down scores across problem complexity, types, and conceptual areas. This approach, termed Targeted Evaluation, is not possible with current benchmarks.
	\item \textbf{Bias Mitigation}: Unlike many datasets derived from public GitHub repositories, which risk overlapping with LLM training data and introducing evaluation bias, CodeEval is distributed solely via a permanent DOI link (Appendix \ref{app:datasheet}). This eliminates the need for expensive data decontamination procedures \cite{li2023textbooks, gunasekar2023textbooks}.
\end{enumerate}
\vspace{-0.5em}
With these features, CodeEval sets a new standard for code generation benchmarks by offering a comprehensive and unbiased framework for targeted evaluation of Large Language Models. A summary of existing benchmark datasets and their comparison with CodeEval is shown in Table \ref{tab:benchmark-datasets}.

\section{Methods}
\label{headings}
\subsection{CodeEval}
\begin{figure*}
    \centering
    \includegraphics[scale=0.3]{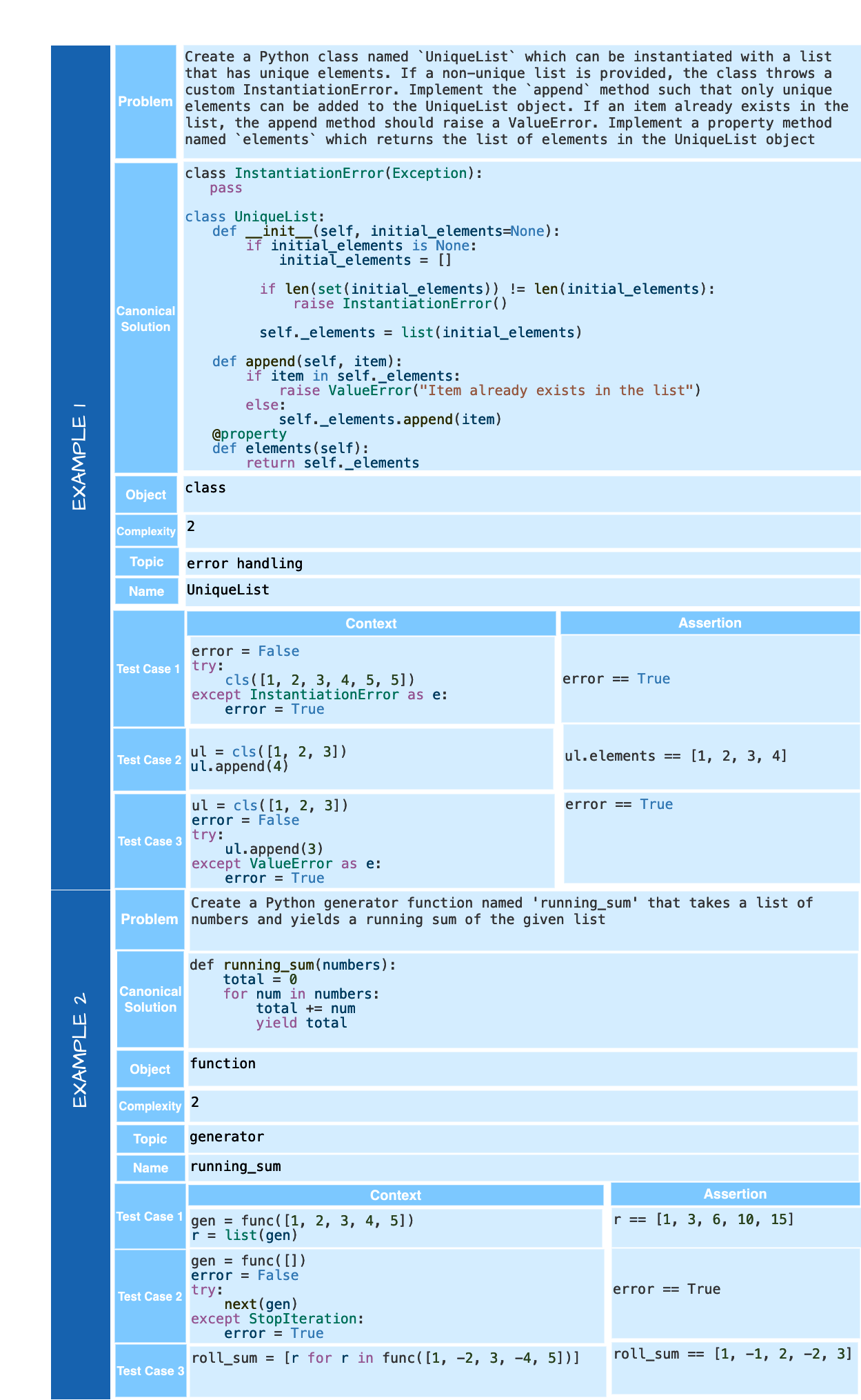}
    \caption{Exemplary problems from CodeEval benchmark dataset. The first problem is of type class and it has three test cases each with a context-assertion pair. The second problem is of type function with three test cases each with a context-assertion pair. Note the usages of cls or func which point to the respective class or function entry-points. The examples also show how context-aware test cases allow for complex definitions of test cases supporting both function and class based problems.}
    \label{fig:problems}           
\end{figure*}
CodeEval comprises 602 hand-crafted problems across 24 Python programming categories (Table \ref{tab:category_definitions}) based on pedagogical foundations \cite{fluentpython} and supplemented with the authors' professional expertise in Python. Problems span three complexity levels—\textit{beginner}, \textit{intermediate}, and \textit{advanced}—validated statistically by correlating with model performance data. The dataset includes both \textit{function} and \textit{class} problem types, enabling multi-dimensional evaluation of coding capabilities. Each problem includes canonical solutions and rigorous test cases. Example problems are shown in Figure \ref{fig:problems} and dataset statistics are provided in Table \ref{tab:dataset_overview}.

\paragraph{Curator Qualifications}
The 602 problems were hand-crafted by the authors, who possess extensive qualifications: (1) PhD-level Computer Science expertise with specialization in software engineering and programming languages, (2) 10-20 years of professional software development experience, particularly in Python programming, (3) Academic research and teaching experience in programming courses and software engineering, especially in Python. This combination of deep theoretical knowledge and substantial practical experience ensures both pedagogical validity and real-world relevance. Each problem underwent multiple rounds of internal review for correctness, clarity, and educational value.

\paragraph{Complexity Level Validation}
  To ensure the robustness and reproducibility of our complexity categorization, we employed a systematic validation approach based on statistical analysis of empirical model performance. Each problem's complexity level was assigned based on the conceptual complexity of required Python constructs, drawing from pedagogical principles and academic curricula. We statistically validated these assignments using model performance data from 15 diverse LLMs (Table \ref{tab:complexity_validation}). Our analysis reveals a strong negative correlation between complexity level and model performance ($r = -0.324$, $p < 0.05$ overall; mean individual model correlation $r = -0.829 \pm 0.394$), confirming that our pedagogically-motivated difficulty levels correspond to empirical performance differences. Paired t-tests demonstrate significant performance drops between complexity levels (Level 2→3: $t = 13.45$, $p < 0.001$), with large effect sizes (Cohen's $d = 0.790$ for Level 1→3) indicating substantial practical significance. This dual validation approach—expert judgment confirmed by empirical evidence—ensures that our complexity levels provide reliable and meaningful difficulty discrimination across diverse model capabilities. 

\begin{figure*}[h]
    \centering
    \includegraphics[scale=0.25]{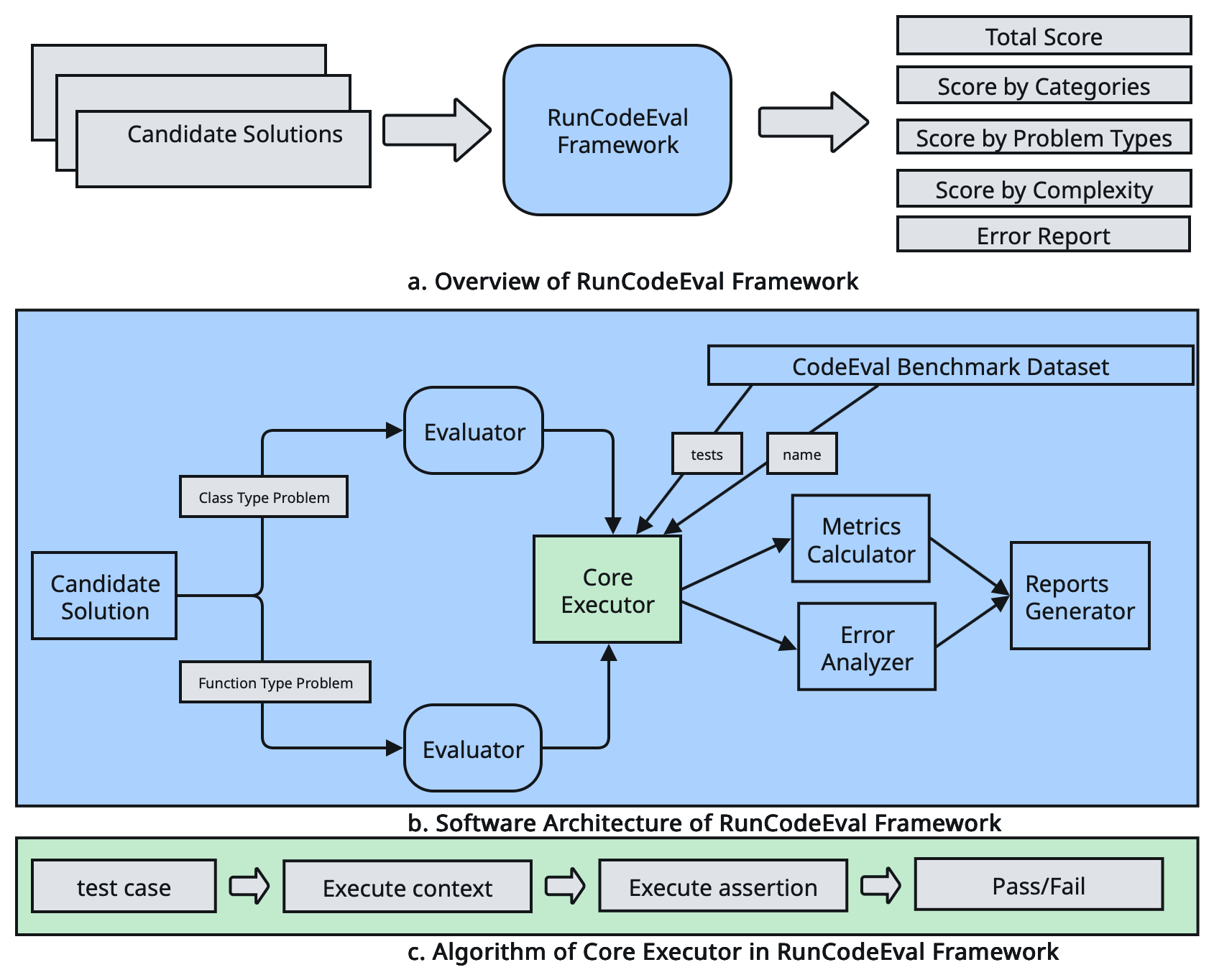}
    \caption{Schematic representation of the RunCodeEval framework. (a) Provides a high-level overview of the framework. (b) Zooms in on the software architecture, detailing its key components. (c) Further focuses on the test execution pipeline, a critical element of the RunCodeEval framework.}      
    \label{fig:runcodeeval}
\end{figure*}

\paragraph{Context-aware Test Cases}
To address the challenge of comprehensive testing, particularly for class-based problems, we introduced context-aware test cases. As shown in Figure \ref{fig:problems}, each test case pairs an optional context with an assertion. This enables sophisticated test scenarios—for instance, Test Case 2 in Example 2 tests exception handling by setting up an empty list context before asserting a StopIteration error, which traditional one-line assertions cannot capture. The context-aware test cases in CodeEval have achieved an average test coverage of $99.1\%$. A more detailed test coverage information is provided in Table \ref{tab:test_coverage} of Appendix \ref{app:dataset_details}.

For simpler test cases that involved basic Python types such as \textit{int}, \textit{float}, \textit{bool}, \textit{str}, \textit{list}, \textit{tuple}, \textit{set}, and \textit{dict}, we used the type-aware mutation technique to extend such test cases to address the test inadequacy problem pointed out by \cite{liu2023your}.

\subsection{RunCodeEval}
The intricate nature of context-aware test cases in CodeEval presents a usability challenge for developing an evaluation pipeline. To facilitate broader adoption and provide a standard operational implementation, we developed RunCodeEval, an open-source execution framework that offers a ready-to-use evaluation solution for CodeEval. RunCodeEval demonstrates how to effectively utilize CodeEval's context-aware test cases and provides validated implementations for test execution, metrics computation, and comprehensive reporting. Figure \ref{fig:runcodeeval} provides a detailed breakdown of the framework's architecture and workflow.

\paragraph{High-Level Overview}

As shown in Figure \ref{fig:runcodeeval}, RunCodeEval serves as the execution engine for the CodeEval benchmark. It processes input files in JSONL format containing LLM-generated solutions, executes these solutions against CodeEval's comprehensive test suite, and generates detailed evaluation reports. This automated pipeline transforms CodeEval's raw problems and test cases into actionable performance insights, including:
\begin{enumerate}[noitemsep, topsep=0pt]
	\item Total scores across all problems,
	\item Scores by complexity levels,
	\item Scores by problem types,
	\item Scores by categories, and
	\item Detailed error reporting, covering issues such as \textit{NameError}, \textit{SyntaxError}, \textit{TimeoutError}, \textit{NoCompletionError}, and general \textit{Error}, whose definitions are provided in Table \ref{tab:benchmark_error_defn}.
\end{enumerate}

\paragraph{Software Architecture}

RunCodeEval operationalizes the CodeEval benchmark by executing LLM solutions against the dataset's comprehensive test suite. Each problem in CodeEval includes multiple test cases with optional context and assertions (see Figure \ref{fig:problems}). RunCodeEval's execution engine transforms these static test definitions into dynamic evaluation processes. As illustrated in Figure \ref{fig:runcodeeval}b, the framework follows these steps:
\begin{enumerate}[noitemsep, topsep=0pt]
	\item Problem Type Identification: The system first determines whether the problem requires a function-level or class-level implementation.
	\item Direct Execution: The evaluator creates an isolated execution namespace and directly executes the candidate solution code within a controlled environment.
	\item Test Execution and Scoring: For each test case, the system executes the optional context setup, runs the assertion against the solution, and tracks pass/fail results. The framework computes the functional correctness score based on the percentage of passed tests.
\end{enumerate}

\paragraph{Test Execution and Partial Credit Scoring}

The execution logic for running test cases is implemented through a direct Python execution approach. As outlined in Figure \ref{fig:runcodeeval}, for each test case, the system:
\begin{enumerate}[noitemsep, topsep=0pt]
	\item Creates an isolated execution namespace to prevent interference between tests,
	\item Executes the context setup code (if defined in the CodeEval dataset),
	\item Runs the assertion against the solution implementation,
	\item Records a pass/fail result with detailed error information when failures occur.
\end{enumerate}
RunCodeEval is equipped to assign partial credit to solutions that are not fully correct. It calculates the functional correctness score by determining the percentage of successfully passed test cases for a given problem. For example, if a solution passes 7 out of 10 test cases, the resulting score is 0.7. The framework includes timeout protection (60 seconds per solution) to handle infinite loops and resource-intensive code. Once all solutions in the dataset are evaluated, a final comprehensive report is generated summarizing the results across multiple dimensions including category, complexity level, and problem type.

\section{Result}
\label{sec:result}
To demonstrate the efficacy of our benchmarking approach, we used RunCodeEval to execute LLM solutions against the comprehensive CodeEval test suite, evaluating 15 popular large language models (Table \ref{tab:model_information}) whose scores are shown in Table~\ref{tab:comprehensive_performance}. RunCodeEval processed CodeEval's 602 programming problems across 24 distinct Python programming categories, executing test cases and computing performance metrics across three complexity levels.

\subsection{Overall Performance}

The evaluated models demonstrated a wide range of capabilities, with overall scores ranging from 21.1\% to 90.5\%. The top-performing model, GPT-4.1 (2025-04-14), achieved 90.5\% accuracy ($\pm$ 2.1\%), while the lowest-performing model, Command-R-Plus (08-2024), achieved only 21.1\% ($\pm$ 3.2\%). This 69.4 percentage point spread highlights the discriminative power of the CodeEval benchmark in differentiating model capabilities.

Based on performance patterns, we identified three distinct tiers of models:
\begin{itemize}[noitemsep, topsep=0pt]
    \item \textbf{Tier 1 (>88\% accuracy):} GPT-4.1, O3-Mini, Qwen3-235B, and O4-Mini, characterized by strong performance across all categories with confidence intervals indicating robust and reliable performance.
    \item \textbf{Tier 2 (82-87\% accuracy):} Claude Sonnet-4, Meta Llama-4-Maverick, Command-A, and Claude Opus-4, showing good overall performance with some category-specific weaknesses.
    \item \textbf{Tier 3 (<82\% accuracy):} The remaining models, exhibiting significant performance gaps and higher variability across problem categories.
\end{itemize}

\subsection{Performance by Complexity Level}

The CodeEval dataset's three complexity levels effectively differentiated model capabilities. As shown in Table~\ref{tab:comprehensive_performance}, all models exhibited consistent performance degradation as complexity increased:

\begin{itemize}[noitemsep, topsep=0pt]
    \item \textbf{Level 1 (Beginner):} Models achieved 89-94\% accuracy, with Meta Llama-4-Scout achieving the highest score at 94.4\% ($\pm$ 3.2\%). The narrow performance range at this level suggests that most modern LLMs handle basic Python constructs competently.
    
    \item \textbf{Level 2 (Intermediate):} Performance ranged from 61.0\% to 94.7\%, with GPT-4.1 leading at 94.7\% ($\pm$ 2.4\%). This level showed clear differentiation between model tiers, with Tier 1 models maintaining >90\% accuracy while lower-tier models dropped below 85\%.
    
    \item \textbf{Level 3 (Advanced):} The most discriminative level, with scores ranging from 20.3\% to 82.8\%. GPT-4.1 maintained the lead with 82.8\% ($\pm$ 4.8\%), while even strong models showed significant performance drops. The average performance decrease from Level 1 to Level 3 was approximately 15-20 percentage points across all models.
\end{itemize}

\begin{table}[htbp]
\centering
\scalebox{0.9}{
\begin{tabular}{p{3.5cm}cc}
\toprule
\textbf{Analysis} & \textbf{Statistic} & \textbf{Value} \\
\midrule
Overall Complexity-Performance Correlation & $r$ & $-0.324$ \\
Correlation Significance & $p$ & $0.025$ \\
Mean Individual Model Correlation & $\bar{r}$ & $-0.829 \pm 0.394$ \\
\midrule
Level 1 vs 2 Difference & $t$ & $1.833$ \\
Level 1 vs 2 Significance & $p$ & $0.067$ \\
Level 2 vs 3 Difference & $t$ & $13.448$ \\
Level 2 vs 3 Significance & $p$ & $< 0.001$ \\
\midrule
Effect Size (L1 vs L2) & $d$ & $0.150$ \\
Effect Size (L2 vs L3) & $d$ & $0.722$ \\
Effect Size (L1 vs L3) & $d$ & $0.790$ \\
\bottomrule
\end{tabular}}
\caption{Statistical Validation of Complexity Levels. $r$ = Pearson correlation coefficient, $t$ = t-statistic, $p$ = significance level, $d$ = Cohen's d effect size.}
\label{tab:complexity_validation}
\end{table}

\subsection{Performance by Problem Type}

The dataset's division between function-based and class-based problems revealed interesting patterns in model capabilities:

\begin{itemize}[noitemsep, topsep=0pt]
    \item \textbf{Function Problems:} Models generally performed better on function-based problems, with scores ranging from 3.3\% to 91.6\%. The top performers (O4-Mini at 91.6\% $\pm$ 2.5\% and GPT-4.1 at 91.4\% $\pm$ 2.5\%) demonstrated strong procedural programming capabilities.
    
    \item \textbf{Class Problems:} Object-oriented problems proved more challenging, with scores ranging from 50.2\% to 89.0\%. GPT-4.1 led with 89.0\% ($\pm$ 3.8\%), but most models showed a 5-10 percentage point drop compared to their function problem performance. Notably, Command-R-08-2024 exhibited an unusual pattern, performing better on class problems (65.9\% $\pm$ 5.9\%) than function problems (47.5\% $\pm$ 4.9\%).
\end{itemize}

\subsection{Category-Specific Performance Analysis}

Analysis of performance across the 24 programming categories revealed both universal strengths and weaknesses among the evaluated models. Comprehensive category-level results for all 15 models are provided in Tables~\ref{tab:performance_by_category_part1} and~\ref{tab:performance_by_category_part2}. Key findings include: (1) fundamental concepts like random module operations and primitive data types achieved consistently high performance (>95\% average) across all models; (2) advanced features such as concurrency and design patterns emerged as the most challenging categories (60-70\% average), effectively differentiating model capabilities. Detailed category-by-category analysis is provided in Appendix~\ref{app:categories}.

\subsection{Error Analysis}

Beyond correctness scores, we analyzed the types and patterns of errors produced by each model. The error analysis revealed significant variation in failure modes, with error rates ranging from 6.0\% (GPT-4.1) to 76.1\% (Command-R-Plus). Runtime and logic errors dominated (12.4\% average), while syntax errors remained relatively rare (1.8\% average), suggesting that modern LLMs have largely mastered Python syntax but struggle with semantic correctness. Notably, error rates showed strong positive correlation with problem complexity, validating the benchmark's difficulty progression. Comprehensive error analysis including breakdown by error type, complexity level, and problem type is presented in Appendix~\ref{app:error_analysis}.

\subsection{Statistical Reliability}

All results are reported with 95\% confidence intervals calculated using the normal approximation for proportions. The consistent sample size of 602 problems per model ensures statistical significance for all reported differences. The narrow confidence intervals (typically $\pm$ 2-6\%) indicate reliable performance estimates, with wider intervals only appearing for the most challenging problem subsets where variance naturally increases.

\subsection{Key Findings}

Our evaluation reveals several important insights:

\begin{enumerate}[noitemsep, topsep=0pt]
    \item \textbf{Clear Performance Stratification:} The CodeEval benchmark successfully differentiates models into distinct performance tiers, providing a reliable metric for comparing code generation capabilities.
    
    \item \textbf{Complexity Sensitivity:} The three-level complexity system effectively captures the degradation in model performance as problems become more challenging, with Level 3 problems serving as particularly strong discriminators.
    
    \item \textbf{Systematic Weaknesses:} All models, regardless of overall performance, struggled with concurrency and advanced object-oriented concepts, suggesting areas for improvement in code-focused language model training.
    
    \item \textbf{Robust Evaluation:} The combination of CodeEval's high problem count (602), diverse categories (24), and multiple complexity levels, executed through RunCodeEval's comprehensive analysis pipeline, provides a robust assessment of code generation capabilities that goes beyond simple accuracy metrics.
\end{enumerate}

These results demonstrate that CodeEval provides a rigorous, pedagogically-grounded benchmark for evaluating code generation capabilities in large language models, with sufficient granularity to identify specific strengths and weaknesses across different programming concepts and complexity levels. RunCodeEval facilitates this evaluation by providing a ready-to-use execution framework whose implementation is non-trivial given the complexity of context-aware test cases in CodeEval.

\subsection{Reproducibility}

To ensure full reproducibility, we provide comprehensive documentation for reproducing both evaluation results using RunCodeEval and LLM solution generation using documented model specifications, prompting strategies, and generation parameters (Appendix~\ref{apx:reproducibility}).

\section{Conclusion}
We presented CodeEval, a pedagogically-grounded benchmark dataset of 602 hand-curated high quality Python problems spanning 24 programming categories with comprehensive test suites (99.1\% coverage). CodeEval's detailed problem specifications and context-aware test cases enable researchers to conduct targeted evaluation across three complexity levels, two problem types and 24 different categories. To facilitate adoption, we also developed RunCodeEval, an execution framework that provides a complete evaluation pipeline for CodeEval, automatically generating fine-grained performance metrics with partial credit scoring. Our evaluation of 15 state-of-the-art LLMs revealed consistent performance degradation with increasing complexity (validated statistically, Cohen's d = 0.790) and universal struggles with advanced concepts like concurrency. CodeEval's comprehensive design enables researchers to gain actionable insights for targeted improvements in LLM code generation capabilities.

\section{Limitations}
\label{sec:limitations}
CodeEval problems are designed to be self-contained, and when dependencies are required, they are limited to Python's standard library. Third-party libraries are intentionally excluded to ensure a focus on fundamental language constructs and algorithmic reasoning, rather than ecosystem-specific tooling. This design choice enables controlled and consistent evaluation across models, which is central to our evaluation goals.

While CodeEval does not assess model performance on scenarios involving popular frameworks (e.g., NumPy, Django, TensorFlow) or complex dependency management, we do not believe this limits real-world relevance. Many developers and researchers care deeply about how well LLMs understand foundational programming semantics, as this forms the bedrock of reliable system behavior. Our evaluation captures this effectively, as demonstrated by the wide performance range across models (90.5\% to 21.1\%) and strong complexity-based separation (Cohen's d = 0.790). This design choice prioritizes fundamental programming concepts and algorithmic thinking, which are essential for assessing core code generation capabilities. 

\bibliography{main}
\bibliographystyle{acl_natbib}

\begin{center}
{\Large \textbf{Appendices}}
\end{center}

\appendix
\section{Dataset Composition and Quality}
\label{app:dataset_details}

This appendix provides comprehensive details about the CodeEval dataset composition, category definitions, and quality metrics that supplement the main dataset description.

\subsection{Dataset Statistics and Distribution}
\begin{table}[t]
\centering
\begin{tabular}{lr}
\toprule
\textbf{Category} & \textbf{Problems} \\
\midrule
operator overloading & 48 (8.0\%) \\
pythonic classes & 36 (6.0\%) \\
generators\_and\_iterators & 34 (5.6\%) \\
design\_pattern & 33 (5.5\%) \\
object identity & 29 (4.8\%) \\
compound data types & 27 (4.5\%) \\
composition & 27 (4.5\%) \\
http\_web & 26 (4.3\%) \\
datetime\_module & 24 (4.0\%) \\
dataclass & 24 (4.0\%) \\
functional programming & 24 (4.0\%) \\
primitive data types & 23 (3.8\%) \\
concurrency & 22 (3.7\%) \\
data structure & 22 (3.7\%) \\
logging\_module & 21 (3.5\%) \\
random module & 21 (3.5\%) \\
decorator & 21 (3.5\%) \\
function argument & 20 (3.3\%) \\
pathlib\_module & 20 (3.3\%) \\
inheritance & 20 (3.3\%) \\
typing & 20 (3.3\%) \\
sorting and slicing & 20 (3.3\%) \\
file\_formats & 20 (3.3\%) \\
error handling & 20 (3.3\%) \\
\bottomrule
\end{tabular}
\caption{Problem Distribution by Category in CodeEval}
\label{tab:category_distribution}
\end{table}

Table~\ref{tab:dataset_overview} presents the statistical overview of the CodeEval dataset, including basic metrics, problem type distribution, and complexity level distribution. For detailed category-level statistics, Table~\ref{tab:category_distribution} shows the complete breakdown of problems across all 24 programming categories. The dataset demonstrates balanced representation with category sizes ranging from 20 to 48 problems, ensuring sufficient statistical power for meaningful evaluation.

\subsection{Category Definitions and Scope}
\begin{table*}[ht]
\centering
\renewcommand{\arraystretch}{0.9} 
\setlength{\tabcolsep}{2pt}  
\begin{adjustbox}{max width=\textwidth} 
\begin{tabular}{p{4cm}p{10cm}}
\toprule
\textbf{Problem Categories} & \textbf{Concept Areas} \\
\midrule
composition & Object composition and delegation patterns \\
compound data types & Dictionary, List, Set, and Tuple operations \\
concurrency & Multi-threading and multi-processing \\
data structure & Selection of efficient data structures for different problem types \\
dataclass & Creation and usage of Python dataclasses \\
datetime module & Date and time manipulation using datetime module \\
decorator & Function and class decorators \\
design pattern & Factory method, dependency injection, singleton, and other design patterns \\
error handling & Error handling with try, except, else, and finally \\
file formats & Reading and writing various file formats (JSON, CSV, XML, etc.) \\
function argument & Passing arguments in Python functions (*args, **kwargs) \\
functional programming & Lambda functions, functools, map, filter, and reduce \\
generators and iterators & Creation and usage of generators and iterators \\
http web & HTTP requests, web APIs, and URL handling \\
inheritance & Inheritance in Object-Oriented Programming \\
logging module & Logging configuration and usage \\
object identity & Object references, deep/shallow copy, and identity \\
operator overloading & Defining and using magic methods for operators \\
pathlib module & File system path operations using pathlib \\
primitive data types & String, float, integer, and boolean data types \\
pythonic classes & Pythonic class design patterns and special methods \\
random module & Random number generation and sampling \\
sorting and slicing & Sorting and slicing for efficient data retrieval \\
typing & Type hints, annotations, and duck typing \\
\bottomrule
\end{tabular}
\end{adjustbox}
\caption{Categories of problems in CodeEval and their evaluated concept areas in Python programming language.}
\label{tab:category_definitions}
\end{table*}
Table~\ref{tab:category_definitions} provides detailed definitions for all 24 programming categories evaluated in CodeEval. Each category represents a distinct area of Python programming knowledge, from fundamental concepts like primitive data types to advanced features like concurrency and design patterns. These categories were selected based on pedagogical importance.

\subsection{Test Coverage Quality Metrics}
\begin{table}[ht]
\centering
\scalebox{0.80}{
\begin{tabular}{lr}
\toprule
\textbf{Characteristic} & \textbf{Value} \\
\midrule
\multicolumn{2}{c}{Basic Metrics} \\
\midrule
Average Test Coverage & 99.1\% \\
Coverage Range & 71.4\% -- 100.0\% \\
\midrule
\multicolumn{2}{c}{Coverage Distribution} \\
\midrule
Excellent ($\geq$90\%) & 577 (95.8\%) \\
Good (70--89\%) & 25 (4.2\%) \\
Fair (50--69\%) & 0 (0.0\%) \\
Poor ($<$50\%) & 0 (0.0\%) \\
\midrule
\multicolumn{2}{c}{Coverage by Category} \\
\midrule
composition & 100.0\% \\
compound\_data\_types & 100.0\% \\
dataclass & 100.0\% \\
decorator & 100.0\% \\
function\_argument & 100.0\% \\
functional\_programming & 100.0\% \\
object\_identity & 100.0\% \\
random\_module & 100.0\% \\
typing & 100.0\% \\
sorting\_and\_slicing & 99.6\% \\
pythonic\_classes & 99.4\% \\
generators\_and\_iterators & 99.3\% \\
error\_handling & 99.2\% \\
design\_pattern & 99.0\% \\
primitive\_data\_types & 98.9\% \\
concurrency & 98.8\% \\
data\_structure & 98.8\% \\
datetime\_module & 98.8\% \\
logging\_module & 98.5\% \\
file\_formats & 98.3\% \\
http\_web & 98.3\% \\
pathlib\_module & 97.9\% \\
operator\_overloading & 97.6\% \\
inheritance & 96.4\% \\
\bottomrule
\end{tabular}}
\caption{Test Coverage Statistics of CodeEval's Test Suite}
\label{tab:test_coverage}
\end{table}

Table~\ref{tab:test_coverage} presents comprehensive test coverage statistics for the CodeEval dataset. The analysis shows exceptionally high coverage, with an average of 99.1\% across all 602 problems. This high coverage ensures that the evaluation accurately captures model performance on the intended programming concepts. The coverage analysis includes both overall statistics and category-specific breakdowns, demonstrating consistent quality across all programming categories.

\section{Performance by Problem Categories}
\label{app:categories}
\begin{table*}[htbp]
\footnotesize
\setlength{\tabcolsep}{4pt}
\scalebox{0.85}{
\begin{tabular}{lp{1.6cm}p{1.6cm}p{1.6cm}p{1.6cm}p{1.6cm}p{1.6cm}p{1.6cm}p{1.6cm}}
\toprule
\textbf{Category} & \textbf{gpt-4.1-2025-04-14} & \textbf{o3-mini-2025-01-31} & \textbf{Qwen3-235B} & \textbf{o4-mini-2025-04-16} & \textbf{sonnet-4} & \textbf{Llama-4-Maverick-17B} & \textbf{command-a-03-2025} & \textbf{opus-4} \\
\midrule
composition & $92.6\pm10.1$ & $92.6\pm10.1$ & $85.2\pm13.7$ & $70.4\pm17.6$ & $92.6\pm10.1$ & $74.1\pm16.8$ & $88.9\pm12.1$ & $63.0\pm18.6$ \\
compound data types & $87.0\pm12.4$ & $88.9\pm10.9$ & $88.9\pm10.9$ & $88.9\pm12.1$ & $83.3\pm13.8$ & $85.2\pm13.7$ & $80.9\pm13.2$ & $87.0\pm12.4$ \\
concurrency & $73.9\pm16.6$ & $77.3\pm16.7$ & $73.9\pm17.5$ & $75.0\pm16.8$ & $65.2\pm18.9$ & $67.8\pm17.4$ & $65.9\pm17.8$ & $60.6\pm17.0$ \\
data structure & $88.6\pm12.8$ & $88.6\pm12.8$ & $88.6\pm12.8$ & $90.9\pm12.3$ & $88.6\pm12.8$ & $95.5\pm6.1$ & $88.6\pm11.0$ & $90.9\pm12.3$ \\
dataclass & $93.8\pm9.0$ & $87.5\pm13.5$ & $91.7\pm11.3$ & $87.5\pm12.2$ & $91.7\pm9.6$ & $79.2\pm16.6$ & $89.6\pm11.8$ & $91.7\pm9.6$ \\
datetime\_module & $98.6\pm2.7$ & $94.4\pm8.5$ & $97.2\pm5.4$ & $95.8\pm4.5$ & $83.3\pm15.2$ & $91.7\pm11.3$ & $97.6\pm4.7$ & $83.3\pm15.2$ \\
decorator & $88.1\pm13.4$ & $88.1\pm13.4$ & $84.1\pm13.8$ & $88.1\pm13.4$ & $72.2\pm18.5$ & $86.5\pm13.5$ & $84.9\pm14.2$ & $81.0\pm15.8$ \\
design\_pattern & $78.8\pm13.5$ & $81.8\pm12.7$ & $69.7\pm15.3$ & $80.3\pm13.4$ & $75.8\pm14.2$ & $73.7\pm14.4$ & $71.2\pm14.8$ & $67.7\pm15.4$ \\
error handling & $84.2\pm11.0$ & $90.0\pm9.0$ & $86.7\pm10.5$ & $86.7\pm10.5$ & $90.0\pm9.0$ & $91.7\pm9.0$ & $86.7\pm10.5$ & $86.7\pm10.5$ \\
file\_formats & $97.5\pm4.9$ & $92.5\pm8.0$ & $97.5\pm4.9$ & $97.5\pm4.9$ & $95.0\pm9.8$ & $92.5\pm10.7$ & $100.0\pm0.0$ & $92.5\pm10.7$ \\
function argument & $95.8\pm5.7$ & $90.8\pm11.0$ & $91.7\pm9.0$ & $90.8\pm11.0$ & $91.7\pm9.0$ & $71.3\pm16.8$ & $79.2\pm15.3$ & $81.7\pm15.2$ \\
functional programming & $93.8\pm9.0$ & $93.8\pm9.0$ & $97.9\pm4.1$ & $93.8\pm9.0$ & $93.8\pm9.0$ & $93.8\pm9.0$ & $88.9\pm10.3$ & $93.8\pm9.0$ \\
generators\_and\_iterators & $94.6\pm6.6$ & $91.2\pm8.7$ & $89.7\pm9.9$ & $91.2\pm8.7$ & $91.2\pm8.7$ & $94.1\pm6.9$ & $97.5\pm3.4$ & $97.1\pm4.0$ \\
http\_web & $95.6\pm5.2$ & $99.2\pm1.5$ & $98.0\pm2.9$ & $99.2\pm1.5$ & $89.2\pm11.1$ & $86.2\pm12.6$ & $89.0\pm11.3$ & $86.2\pm13.0$ \\
inheritance & $90.0\pm11.5$ & $85.0\pm14.4$ & $85.0\pm14.4$ & $85.0\pm14.4$ & $87.5\pm12.1$ & $92.5\pm10.7$ & $80.0\pm16.5$ & $85.0\pm12.5$ \\
logging\_module & $79.4\pm14.9$ & $75.6\pm15.2$ & $79.4\pm13.0$ & $85.2\pm11.1$ & $70.5\pm19.6$ & $89.4\pm10.5$ & $82.2\pm15.2$ & $75.2\pm18.5$ \\
object identity & $94.2\pm8.0$ & $96.5\pm6.8$ & $90.8\pm10.2$ & $96.5\pm6.8$ & $94.2\pm8.0$ & $90.8\pm10.2$ & $82.8\pm14.0$ & $94.2\pm8.0$ \\
operator overloading & $81.2\pm11.2$ & $72.9\pm12.7$ & $82.3\pm10.7$ & $65.6\pm13.4$ & $78.1\pm11.6$ & $74.0\pm12.4$ & $70.1\pm12.6$ & $78.1\pm11.6$ \\
pathlib\_module & $86.7\pm12.9$ & $86.7\pm12.9$ & $81.7\pm13.0$ & $86.7\pm12.0$ & $75.0\pm18.3$ & $81.7\pm13.0$ & $73.3\pm16.8$ & $68.3\pm19.8$ \\
primitive data types & $95.7\pm5.9$ & $93.5\pm9.4$ & $95.7\pm5.9$ & $97.8\pm4.3$ & $97.8\pm4.3$ & $97.8\pm4.3$ & $95.7\pm5.9$ & $97.8\pm4.3$ \\
pythonic classes & $95.8\pm6.0$ & $95.8\pm6.0$ & $93.1\pm8.0$ & $90.3\pm9.4$ & $91.7\pm9.2$ & $89.6\pm9.4$ & $94.4\pm7.6$ & $87.5\pm9.9$ \\
random module & $100.0\pm0.0$ & $100.0\pm0.0$ & $100.0\pm0.0$ & $100.0\pm0.0$ & $100.0\pm0.0$ & $100.0\pm0.0$ & $95.2\pm9.3$ & $100.0\pm0.0$ \\
sorting and slicing & $100.0\pm0.0$ & $100.0\pm0.0$ & $100.0\pm0.0$ & $100.0\pm0.0$ & $100.0\pm0.0$ & $98.8\pm2.5$ & $98.8\pm2.5$ & $100.0\pm0.0$ \\
typing & $95.0\pm6.7$ & $98.8\pm2.5$ & $98.3\pm3.3$ & $97.5\pm4.9$ & $92.5\pm10.7$ & $98.3\pm3.3$ & $95.0\pm6.7$ & $92.5\pm10.7$ \\
\bottomrule
\end{tabular}}
\caption{Model Performance by Programming Category for Top-Performing Models. Results show accuracy percentages with 95\% confidence intervals across 24 Python programming categories for the highest-performing 8 models. See Appendix \ref{app:model_info} for model information.}
\label{tab:performance_by_category_part1}
\end{table*}

This appendix provides detailed analysis of model performance across all 24 programming categories in the CodeEval dataset. The comprehensive results are presented in Tables~\ref{tab:performance_by_category_part1} and~\ref{tab:performance_by_category_part2}, which together cover all 15 evaluated models across the complete category spectrum.

\subsection{Performance Patterns by Category}
\begin{table*}[htbp]
\footnotesize
\scalebox{0.9}{
\begin{tabular}{lp{1.6cm}p{1.6cm}p{1.6cm}p{1.6cm}p{1.6cm}p{1.6cm}p{1.6cm}}
\toprule
\textbf{Category} & \textbf{grok-2-vision-1212} & \textbf{Llama-4-Scout-17B} & \textbf{gemini-2.0-flash} & \textbf{gpt-3.5-turbo} & \textbf{Qwen3-4B} & \textbf{command-r-08-2024} & \textbf{command-r-plus-08-2024} \\
\midrule
composition & $59.3\pm18.9$ & $59.3\pm18.9$ & $77.8\pm16.0$ & $88.9\pm12.1$ & $51.9\pm19.2$ & $74.1\pm16.8$ & $44.4\pm19.1$ \\
compound data types & $83.3\pm13.8$ & $85.2\pm12.6$ & $87.0\pm12.4$ & $79.0\pm14.3$ & $79.6\pm15.0$ & $40.7\pm18.9$ & $0.0\pm0.0$ \\
concurrency & $65.9\pm19.8$ & $60.2\pm17.6$ & $60.6\pm20.5$ & $67.4\pm16.7$ & $35.2\pm15.6$ & $33.0\pm19.2$ & $4.5\pm8.9$ \\
data structure & $84.1\pm13.5$ & $90.9\pm10.5$ & $86.4\pm13.2$ & $84.1\pm13.5$ & $88.6\pm11.0$ & $25.0\pm18.0$ & $0.0\pm0.0$ \\
dataclass & $91.0\pm9.4$ & $84.7\pm13.6$ & $91.7\pm9.6$ & $81.9\pm14.7$ & $72.9\pm17.7$ & $65.3\pm18.1$ & $41.7\pm20.1$ \\
datetime\_module & $83.3\pm15.2$ & $92.0\pm10.8$ & $72.2\pm16.5$ & $76.7\pm16.4$ & $71.2\pm17.5$ & $12.8\pm13.5$ & $0.0\pm0.0$ \\
decorator & $69.0\pm19.7$ & $81.8\pm15.6$ & $61.1\pm20.3$ & $65.9\pm19.7$ & $63.5\pm20.6$ & $19.1\pm17.2$ & $0.0\pm0.0$ \\
design\_pattern & $67.2\pm15.1$ & $60.6\pm16.4$ & $69.7\pm15.3$ & $63.6\pm16.1$ & $56.6\pm16.4$ & $37.4\pm16.0$ & $21.2\pm14.2$ \\
error handling & $87.5\pm9.7$ & $89.2\pm9.9$ & $79.2\pm13.6$ & $82.5\pm11.0$ & $79.2\pm13.6$ & $60.8\pm17.0$ & $17.5\pm16.3$ \\
file\_formats & $95.0\pm9.8$ & $82.5\pm16.3$ & $75.0\pm19.5$ & $80.0\pm18.0$ & $77.5\pm16.6$ & $57.5\pm21.7$ & $5.0\pm9.8$ \\
function argument & $91.7\pm9.0$ & $81.7\pm15.2$ & $83.3\pm11.6$ & $85.0\pm12.7$ & $72.5\pm18.2$ & $60.0\pm19.7$ & $5.0\pm9.8$ \\
functional programming & $93.2\pm9.0$ & $92.7\pm9.1$ & $91.0\pm10.2$ & $92.7\pm9.1$ & $89.6\pm10.6$ & $66.7\pm19.3$ & $0.0\pm0.0$ \\
generators\_and\_iterators & $98.5\pm2.9$ & $98.5\pm2.9$ & $91.2\pm8.7$ & $89.7\pm9.1$ & $91.2\pm8.7$ & $57.4\pm16.6$ & $27.9\pm15.0$ \\
http\_web & $90.0\pm9.9$ & $92.3\pm8.7$ & $82.3\pm13.5$ & $85.9\pm13.2$ & $68.5\pm15.5$ & $82.6\pm13.1$ & $0.0\pm0.0$ \\
inheritance & $65.0\pm20.2$ & $82.5\pm14.7$ & $85.0\pm14.4$ & $67.5\pm19.2$ & $62.5\pm20.0$ & $47.5\pm21.9$ & $37.5\pm21.2$ \\
logging\_module & $68.9\pm19.4$ & $82.7\pm13.2$ & $61.0\pm21.0$ & $69.8\pm18.8$ & $62.2\pm19.7$ & $18.1\pm15.3$ & $0.0\pm0.0$ \\
object identity & $90.8\pm10.2$ & $87.4\pm11.9$ & $89.7\pm10.3$ & $82.8\pm14.0$ & $86.2\pm12.8$ & $62.1\pm18.0$ & $34.5\pm17.6$ \\
operator overloading & $74.8\pm11.9$ & $67.5\pm13.1$ & $65.6\pm13.4$ & $61.5\pm13.8$ & $54.5\pm14.0$ & $64.9\pm13.3$ & $69.1\pm12.9$ \\
pathlib\_module & $75.0\pm18.3$ & $71.7\pm18.5$ & $61.7\pm21.3$ & $71.7\pm19.7$ & $56.7\pm21.8$ & $10.0\pm13.5$ & $0.0\pm0.0$ \\
primitive data types & $95.7\pm5.9$ & $97.8\pm4.3$ & $97.8\pm4.3$ & $89.1\pm12.3$ & $93.5\pm9.4$ & $80.4\pm16.0$ & $13.0\pm14.1$ \\
pythonic classes & $88.9\pm9.7$ & $87.5\pm10.6$ & $91.7\pm9.2$ & $90.7\pm8.4$ & $85.8\pm10.9$ & $91.0\pm9.2$ & $75.8\pm13.7$ \\
random module & $95.2\pm9.3$ & $87.3\pm13.9$ & $89.3\pm12.9$ & $90.5\pm12.9$ & $100.0\pm0.0$ & $82.5\pm16.0$ & $0.0\pm0.0$ \\
sorting and slicing & $98.8\pm2.5$ & $96.2\pm5.4$ & $98.8\pm2.5$ & $96.2\pm5.4$ & $100.0\pm0.0$ & $61.2\pm20.9$ & $0.0\pm0.0$ \\
typing & $93.3\pm10.2$ & $93.3\pm10.2$ & $95.0\pm9.8$ & $90.0\pm11.5$ & $92.1\pm7.4$ & $55.0\pm22.4$ & $5.0\pm9.8$ \\
\bottomrule
\end{tabular}}
\caption{Model Performance by Programming Category for Remaining Models. Results show accuracy percentages with 95\% confidence intervals across 24 Python programming categories for the remaining 7 evaluated models. See Appendix \ref{app:model_info} for model information.}
\label{tab:performance_by_category_part2}
\end{table*}

\textbf{Strongest Categories:} Random module operations, sorting and slicing, and primitive data types showed consistently high performance across all models, with many achieving perfect or near-perfect scores. These categories represent fundamental programming concepts that are well-represented in training data. The low coefficient of variation (CV < 10\%) for these categories indicates consistent performance across all evaluated models.

\textbf{Most Challenging Categories:} Concurrency, design patterns, and operator overloading emerged as the most discriminative categories. Even top-tier models showed significant performance drops in these areas, with concurrency problems averaging 60-70\% accuracy across all models. These categories exhibited the highest coefficient of variation (CV > 30\%), effectively differentiating model capabilities. The challenges in these areas require deep understanding of advanced Python concepts and complex programming paradigms.

\subsection{Category-Level Insights}

\textbf{Fundamental Concepts:} Categories like primitive data types, sorting and slicing, and random module operations consistently achieved >90\% accuracy across most models, suggesting these concepts are well-captured in training data.

\textbf{Object-Oriented Programming:} Categories involving inheritance, pythonic classes, and dataclass showed moderate performance variation (70-95\%), with top-tier models demonstrating superior understanding of Python's object-oriented paradigms.

\textbf{Advanced Features:} Concurrency, decorator patterns, and operator overloading proved most challenging, with even GPT-4.1 achieving only 73-88\% accuracy in these areas, highlighting the complexity of advanced Python programming concepts.

\textbf{Standard Library:} Performance on modules like datetime, pathlib, and logging showed wide variation (13-99\%), suggesting that familiarity with specific library APIs varies significantly across models and may depend on training data representation.
\section{Error Analysis}
\label{app:error_analysis}

\begin{table}[ht]
\centering
\renewcommand{\arraystretch}{0.9} 
\setlength{\tabcolsep}{2pt}  
\begin{adjustbox}{max width = \textwidth}
\begin{tabular}{p{0.4\linewidth} p{0.6\linewidth}}
\toprule
\textbf{Error Type} & \textbf{Definition} \\
\midrule
NameError &  The generated code has undefined reference commonly caused by missing import statements \\
SyntaxError &  The generated code has syntax error \\
TimeoutError &  The generated code runtime exceeded 60 seconds\\ 
NoCompletionError & The model did not generate any code in its completion\\
Error & All other errors fall within this category\\
\bottomrule
\end{tabular}
\end{adjustbox}
\caption{Definitions of error types captured by RunCodeEval software framework.}
\label{tab:benchmark_error_defn}
\end{table}
\begin{table*}[htbp]
\centering
\resizebox{\textwidth}{!}{
\begin{tabular}{l|c|ccc|cc}
\toprule
\textbf{Model} & \textbf{Overall} & \multicolumn{3}{c|}{\textbf{Complexity}} & \multicolumn{2}{c}{\textbf{Problem Type}} \\
 & Error Rate & Beginner & Intermediate & Advanced & Function & Class \\
\midrule
gpt-4.1-(04-14) & $6.0\pm1.9$ & $10.4\pm5.3$ & $8.0\pm3.3$ & $22.7\pm5.8$ & $12.8\pm3.4$ & $14.5\pm4.6$ \\
Qwen3-235B & $6.6\pm2.0$ & $10.4\pm5.3$ & $10.6\pm3.8$ & $24.6\pm5.9$ & $14.4\pm3.6$ & $16.7\pm4.9$ \\
o3-mini-(01-31) & $7.1\pm2.1$ & $12.6\pm5.7$ & $9.1\pm3.5$ & $20.7\pm5.6$ & $12.8\pm3.4$ & $15.4\pm4.7$ \\
Llama-4-Maverick-17B & $7.5\pm2.1$ & $9.6\pm5.1$ & $15.5\pm4.4$ & $25.1\pm6.0$ & $15.5\pm3.7$ & $20.6\pm5.3$ \\
o4-mini-(04-16) & $7.8\pm2.2$ & $9.6\pm5.1$ & $10.2\pm3.7$ & $26.1\pm6.1$ & $12.0\pm3.3$ & $21.1\pm5.3$ \\
command-a-03-2025 & $8.8\pm2.3$ & $12.6\pm5.7$ & $15.2\pm4.4$ & $27.6\pm6.2$ & $17.9\pm3.9$ & $20.2\pm5.2$ \\
claude-sonnet-4 & $9.1\pm2.3$ & $8.1\pm4.8$ & $11.0\pm3.8$ & $27.6\pm6.2$ & $14.4\pm3.6$ & $18.4\pm5.1$ \\
Llama-4-Scout-17B & $10.1\pm2.4$ & $9.6\pm5.1$ & $20.1\pm4.9$ & $33.0\pm6.5$ & $18.4\pm3.9$ & $28.1\pm5.8$ \\
grok-2-vision-1212 & $11.3\pm2.5$ & $10.4\pm5.3$ & $18.6\pm4.7$ & $30.0\pm6.3$ & $16.0\pm3.7$ & $28.1\pm5.8$ \\
claude-opus-4 & $11.3\pm2.5$ & $8.9\pm4.9$ & $14.4\pm4.3$ & $31.0\pm6.4$ & $15.2\pm3.7$ & $24.6\pm5.6$ \\
gemini-2.0-flash & $13.6\pm2.7$ & $11.9\pm5.6$ & $19.3\pm4.8$ & $34.5\pm6.5$ & $21.1\pm4.1$ & $25.4\pm5.7$ \\
gpt-3.5-turbo & $14.6\pm2.8$ & $14.8\pm6.1$ & $21.2\pm4.9$ & $35.5\pm6.6$ & $22.7\pm4.3$ & $27.6\pm5.8$ \\
Qwen3-4B & $21.3\pm3.3$ & $14.8\pm6.1$ & $27.7\pm5.4$ & $46.8\pm6.9$ & $27.0\pm4.5$ & $38.2\pm6.3$ \\
command-r-08-2024 & $41.4\pm3.9$ & $43.7\pm8.4$ & $42.8\pm6.0$ & $60.1\pm6.7$ & $55.1\pm5.0$ & $38.6\pm6.3$ \\
command-r-plus-08-2024 & $76.1\pm3.4$ & $87.4\pm5.7$ & $75.0\pm5.2$ & $80.3\pm5.5$ & $96.8\pm1.8$ & $51.3\pm6.5$ \\
\bottomrule
\end{tabular}}
\caption{Comprehensive Error Analysis Across Models. Error rates (\%) with 95\% confidence intervals broken down by overall performance, complexity levels (Beginner, Intermediate, Advanced), and problem types (Function vs Class). See Appendix \ref{app:model_info} for model information.}
\label{tab:comprehensive_errors}
\end{table*}

\begin{table*}[htbp]
\centering
\resizebox{\textwidth}{!}{
\begin{tabular}{l|ccccc}
\toprule
\textbf{Model} & \textbf{SyntaxError} & \textbf{NameError} & \textbf{TimeoutError} & \textbf{NoCompletionError} & \textbf{Other Errors} \\
\midrule
gpt-4.1-(04-14) & $0.00\pm0.32$ & $0.5\pm0.6$ & $0.00\pm0.32$ & $0.00\pm0.32$ & $16.1\pm2.9$ \\
Qwen3-235B & $0.00\pm0.32$ & $2.1\pm1.2$ & $0.00\pm0.32$ & $0.00\pm0.32$ & $14.5\pm2.8$ \\
o3-mini-(01-31) & $0.00\pm0.32$ & $0.4\pm0.6$ & $0.00\pm0.32$ & $0.00\pm0.32$ & $16.2\pm3.0$ \\
Llama-4-Maverick-17B & $2.2\pm1.2$ & $0.00\pm0.32$ & $0.00\pm0.32$ & $0.00\pm0.32$ & $14.4\pm2.8$ \\
o4-mini-(04-16) & $0.00\pm0.32$ & $0.00\pm0.32$ & $0.00\pm0.32$ & $0.4\pm0.6$ & $16.3\pm3.0$ \\
command-a-03-2025 & $0.3\pm0.5$ & $0.3\pm0.5$ & $0.00\pm0.32$ & $0.00\pm0.32$ & $16.0\pm2.9$ \\
claude-sonnet-4 & $5.1\pm1.8$ & $0.6\pm0.7$ & $0.00\pm0.32$ & $0.00\pm0.32$ & $10.9\pm2.5$ \\
Llama-4-Scout-17B & $0.3\pm0.5$ & $0.00\pm0.32$ & $0.00\pm0.32$ & $0.00\pm0.32$ & $16.3\pm3.0$ \\
grok-2-vision-1212 & $3.7\pm1.5$ & $0.5\pm0.6$ & $0.00\pm0.32$ & $0.00\pm0.32$ & $12.5\pm2.6$ \\
claude-opus-4 & $4.9\pm1.7$ & $0.5\pm0.6$ & $0.00\pm0.32$ & $0.00\pm0.32$ & $11.2\pm2.5$ \\
gemini-2.0-flash & $7.3\pm2.1$ & $0.2\pm0.5$ & $0.00\pm0.32$ & $0.00\pm0.32$ & $9.1\pm2.3$ \\
gpt-3.5-turbo & $2.3\pm1.2$ & $0.6\pm0.7$ & $0.00\pm0.32$ & $0.00\pm0.32$ & $13.8\pm2.8$ \\
Qwen3-4B & $1.3\pm1.0$ & $1.0\pm0.9$ & $0.00\pm0.32$ & $0.00\pm0.32$ & $14.3\pm2.8$ \\
command-r-08-2024 & $0.07\pm0.38$ & $0.07\pm0.38$ & $0.00\pm0.32$ & $12.6\pm2.7$ & $3.9\pm1.6$ \\
command-r-plus-08-2024 & $0.00\pm0.32$ & $0.00\pm0.32$ & $0.00\pm0.32$ & $15.8\pm2.9$ & $0.8\pm0.8$ \\
\bottomrule
\end{tabular}}
\caption{Error Type Distribution Across Models. Error rates broken down by specific error types (\%) with 95\% confidence intervals, including SyntaxError, NameError, TimeoutError, NoCompletionError, and Other runtime/logic errors. See Appendix \ref{app:model_info} for model information.}
\label{tab:error_type_breakdown}
\end{table*}

This appendix provides a comprehensive analysis of errors encountered during the evaluation of 15 large language models on the CodeEval benchmark. Tables~\ref{tab:comprehensive_errors} and~\ref{tab:error_type_breakdown} present detailed error statistics across multiple dimensions. The definitions of the error type is shown in Table \ref{tab:benchmark_error_defn}

\subsection{Overall Error Patterns}

The error analysis reveals significant variation in failure rates across models, ranging from 6.0\% (GPT-4.1) to 76.1\% (Command-R-Plus). This 70 percentage point spread in error rates demonstrates the benchmark's ability to differentiate model robustness in addition to correctness. The average error rate across all models was 16.8\%, indicating that while modern LLMs have made substantial progress in code generation, reliable error-free code synthesis remains challenging.

\subsection{Error Analysis by Complexity}

Error rates show a clear progression with problem complexity:

\begin{itemize}
    \item \textbf{Level 1 (Beginner):} Error rates range from 8.1\% to 87.4\%, with most high-performing models maintaining error rates below 15\%. The wide range suggests that even basic problems can expose fundamental limitations in some models.
    
    \item \textbf{Level 2 (Intermediate):} Error rates increase to 8.0\%-75.0\%, with the median around 20\%. The increased variance at this level effectively differentiates model capabilities in handling moderately complex programming tasks.
    
    \item \textbf{Level 3 (Advanced):} Error rates span 20.7\%-80.3\%, with even top-tier models showing error rates above 20\%. This demonstrates the challenge of generating correct code for complex programming scenarios.
\end{itemize}

The consistent increase in error rates with complexity validates the benchmark's difficulty progression and highlights that advanced problems remain challenging even for state-of-the-art models.

\subsection{Error Analysis by Problem Type}

The analysis reveals interesting patterns between function and class-based problems:

\begin{itemize}
    \item \textbf{Function Problems:} Generally show higher error rates (12.0\%-96.8\%), suggesting that functional programming tasks may have more edge cases or require more precise implementations.
    
    \item \textbf{Class Problems:} Display more moderate error rates (14.5\%-51.3\%), though with higher variance. Notably, some models that performed poorly overall (e.g., Command-R models) showed relatively better performance on class problems, suggesting different architectural strengths.
\end{itemize}

\subsection{Error Type Distribution}

Table~\ref{tab:error_type_breakdown} reveals the distribution of specific error types:

\begin{itemize}
    \item \textbf{Runtime/Logic Errors ("Other Errors"):} Dominate the error landscape, accounting for 12.4\% of all failures on average. These errors indicate that models can generate syntactically correct code that fails during execution, highlighting the challenge of semantic correctness.
    
    \item \textbf{Syntax Errors:} Relatively rare (1.8\% average), with some models achieving near-zero syntax error rates. However, certain models (e.g., Gemini-2.0-Flash at 7.3\%) show higher rates, suggesting variations in syntactic understanding.
    
    \item \textbf{NoCompletionError:} Primarily affects Command-R models (12.6\%-15.8\%), indicating generation failures or truncation issues specific to certain model architectures.
    
    \item \textbf{NameError and TimeoutError:} Minimal occurrence (<0.5\% average), suggesting that undefined variable usage and infinite loops are well-handled by most modern LLMs.
\end{itemize}

\subsection{Key Insights from Error Analysis}

\begin{enumerate}
    \item \textbf{Error Rate as a Quality Metric:} Beyond accuracy scores, error rates provide crucial insights into model reliability. The best-performing models maintain low error rates (<10\%) while achieving high accuracy.
    
    \item \textbf{Complexity-Error Correlation:} The strong positive correlation between problem complexity and error rates validates the benchmark's design and suggests that complex problems effectively stress-test model capabilities.
    
    \item \textbf{Model-Specific Patterns:} Different models exhibit characteristic error patterns. For instance, Claude models show higher syntax error rates, while Command-R models struggle with completion, suggesting architectural differences in handling code generation tasks.
    
    \item \textbf{Runtime Errors Predominate:} The dominance of runtime/logic errors over syntax errors indicates that modern LLMs have largely mastered Python syntax but struggle with semantic correctness and edge case handling.
\end{enumerate}

These error patterns provide valuable insights for both model developers and users, highlighting specific areas where current code generation models need improvement and helping users understand the types of failures to expect when deploying these models in practice.

\section{Evaluated Models Information}
\label{app:model_info}

Table~\ref{tab:model_information} provides comprehensive information about all 15 large language models evaluated in this study, including model specifications, providers, and reference URLs for additional details.

\begin{table*}[htbp]
\centering
\footnotesize
\begin{tabular}{p{3.5cm}p{2cm}p{2cm}p{2cm}p{6cm}}
\toprule
\textbf{Model Name} & \textbf{Provider} & \textbf{Release Date} & \textbf{Model Type} & \textbf{Reference} \\
\midrule
gpt-4.1-2025-04-14 & OpenAI & Apr 2025 & Commercial API & \cite{openai2024models} \\
\midrule
o3-mini-2025-01-31 & OpenAI & Jan 2025 & Commercial API & \cite{openai2024models} \\
\midrule
o4-mini-2025-04-16 & OpenAI & Apr 2025 & Commercial API & \cite{openai2024models} \\
\midrule
claude-sonnet-4-20250514 & Anthropic & May 2025 & Commercial API & \cite{anthropic2024claude} \\
\midrule
claude-opus-4-20250514 & Anthropic & May 2025 & Commercial API & \cite{anthropic2024claude} \\
\midrule
gpt-3.5-turbo & OpenAI & Mar 2023 & Commercial API & \cite{openai2024models} \\
\midrule
Qwen/Qwen3-235B-A22B-Instruct-2507 & Alibaba Cloud & Jul 2025 & Open Source & \cite{qwen2024huggingface} \\
\midrule
Qwen/Qwen3-4B & Alibaba Cloud & 2025 & Open Source & \cite{qwen2024huggingface} \\
\midrule
meta-llama/Llama-4-Maverick-17B-128E-Instruct & Meta & 2025 & Open Source & \cite{meta2024llama} \\
\midrule
meta-llama/Llama-4-Scout-17B-16E-Instruct & Meta & 2025 & Open Source & \cite{meta2024llama} \\
\midrule
command-a-03-2025 & Cohere & Mar 2025 & Commercial API & \cite{cohere2024models} \\
\midrule
command-r-08-2024 & Cohere & Aug 2024 & Commercial API & \cite{cohere2024models} \\
\midrule
command-r-plus-08-2024 & Cohere & Aug 2024 & Commercial API & \cite{cohere2024models} \\
\midrule
grok-2-vision-1212 & xAI & Dec 2024 & Commercial API & \cite{xai2024grok} \\
\midrule
gemini-2.0-flash & Google & 2024 & Commercial API & \cite{google2024gemini} \\
\bottomrule
\end{tabular}
\caption{Comprehensive Information for Evaluated Large Language Models. All models were accessed via their respective APIs or platforms during the evaluation period (2024-2025).}
\label{tab:model_information}
\end{table*}

\section{Accessing CodeEval and RunCodeEval}
\label{app:access}
\begin{table*}
    \centering
    \scalebox{0.9}{
    \small
    \begin{tabular}{lcr}
    \toprule
    \textbf{Keys} &  \textbf{Value} & \textbf{Data type} \\
    \midrule
    task\_id & Unique problem identifier & String \\
    problem & Natural language description of the problem & String \\
    topic &  The category of the problem & String\\
    object & The problem type - function or class & String \\
    name & the name of the function or class that is an entry-point to the model generated code & String\\
    canonical\_solution & The canonical solution to the problem & String \\
    tests & the test cases to get functional correctness score & String\\
    complexity & The complexity tier of the problem - 1, 2 or 3 representing beginner, intermediate or advanced & Integer \\
    \bottomrule
    \end{tabular}}
    \caption{Data Format of CodeEval dataset}
    \label{tab:data_format}
\end{table*}
The CodeEval dataset is provided in JSONL format, where each line represents a single dataset instance in JSON structure. The keys, along with their corresponding value definitions and data types, are listed in Table \ref{tab:data_format}. The dataset metadata, available in Croissant format \cite{akhtar2024croissant}, can be accessed at \url{https://github.com/dannybrahman/runcodeeval/blob/main/dataset/croissant.json}. The dataset files can be read using any general-purpose programming language. The RunCodeEval software framework is available in the GitHub repository at \url{https://github.com/dannybrahman/runcodeeval}. While the CodeEval dataset can be directly downloaded from its permanent DOI link at \url{https://doi.org/10.5281/zenodo.17495202}, we also provide a script in the RunCodeEval repository to facilitate programmatic access.

\section{Reproduciblity of the experiments}
\label{apx:reproducibility}
We provide comprehensive reproducibility support for both the evaluation results and the underlying LLM solutions presented in this paper.

\subsection{Reproducing Evaluation Results}

The evaluation results presented in this paper can be reproduced using the RunCodeEval software framework, which is freely and publicly available at Github repository \url{https://github.com/dannybrahman/runcodeeval}. RunCodeEval provides deterministic evaluation with fixed random seeds and consistent test execution environments, ensuring reproducible metrics across different systems.

\subsection{Reproducing LLM Solutions}

To facilitate complete reproducibility, we provide detailed information for regenerating the LLM solutions whose code is also freely and openly available at the RunCodeEval's GitHub repository -- \url{https://github.com/dannybrahman/runcodeeval/tree/main/llm_solutions}:

\begin{itemize}
    \item \textbf{Model Specifications:} All evaluated models are publicly available through their respective APIs or platforms (OpenAI, Anthropic, Google, Cohere, Meta, Mistral, xAI, DeepSeek, Qwen). Model version identifiers are provided in Table~\ref{tab:model_information}.
    
    \item \textbf{Prompting Strategy:} We used a consistent prompting approach across all models, providing the problem description from CodeEval and requesting a complete Python solution. No few-shot examples or chain-of-thought prompting was used to ensure fair comparison.
    
    \item \textbf{Generation Parameters:} For consistent results, we used default values for parameters unless otherwise specified by the model provider. The associated code with this paper (which will also be freely and openly available) has details on generation parameters.
    
    \item \textbf{Solution Collection:} The repository code includes scripts and configurations for collecting solutions from all 15 evaluated models, including API endpoints, authentication setup, rate limiting configurations and generation parameters.
\end{itemize}

\subsection{Reproducibility Considerations}

While we provide comprehensive reproduction guidance, researchers should note that:

\begin{enumerate}
    \item LLM API responses may vary slightly due to model updates or infrastructure changes, even with deterministic settings.
    \item Some models may have been updated or deprecated since our evaluation; we recommend using the specific model versions listed in Table~\ref{tab:model_information}.
    \item API access and pricing may vary by provider and user account status.
\end{enumerate}

Complete reproduction instructions, including detailed setup guides and example commands, are provided in the repository documentation.

\section{CodeEval Datasheet}
\label{app:datasheet}
In this section, we provide datasheet \cite{gebru2021datasheets} of the CodeEval dataset published with a permanent doi \url{https://doi.org/10.5281/zenodo.17495202}
\subsection{Motivation}

\begin{enumerate}
    \item For what purpose was the dataset created? Was there a specific task in mind? Was there a specific gap that needed to be filled? Please provide a description. \\
    \textcolor{blue}{The dataset was created to enable evaluation of code understanding or reasoning capabilities of Large Language Models (LLMs) trained on code. Currently, no benchmark dataset exists that allows for targeted improvements of LLMs - this gap is fulfilled by CodeEval. Moreover, the current benchmark datasets are incomprehensive as they focus on individual aspects of evaluation such as complexity, function-based problems or class-based problems.}
    \item Who created the dataset (e.g., which team, research group) and on
behalf of which entity (e.g., company, institution, organization)? \\
    \textcolor{blue}{The dataset was primarily created by Danny Brahman (PhD student at University of Denver) in collaboration with his advisor Dr. Mohammad Mahoor (Professor of Computer Science Department at University of Denver).}
    \item Who funded the creation of the dataset? If there is an associated
grant, please provide the name of the grantor and the grant name and number \\
    \textcolor{blue}{The work was not funded by any grant.}
    \item Any other comment? \\
    \textcolor{blue}{No.}
\end{enumerate}

\subsection{Composition}

\begin{enumerate}
\item What do the instances that comprise the dataset represent (e.g.,
documents, photos, people, countries)? Are there multiple types of
instances (e.g., movies, users, and ratings; people and interactions between them; nodes and edges)? Please provide a description. \\
\textcolor{blue}{Each instance of the dataset represents a natural language description of a Python problem, the problem's solution code in Python, one of the 24 categories the problem belongs to, the complexity level of problem (beginner, intermediate, advanced), and the test cases for evaluating functional correctness of a model generated code for this problem.}
\item How many instances are there in total (of each type, if appropriate)? \\
\textcolor{blue}{There are a total of 602 instances distributed among 24 categories.}
\item Does the dataset contain all possible instances or is it a sample (not necessarily random) of instances from a larger set? If the dataset is a sample, then what is the larger set? Is the sample representative of the larger set (e.g., geographic coverage)? If so, please describe how this representativeness was validated/verified. If it is not representative of the larger set, please describe why not (e.g., to cover a more diverse range of instances, because instances were withheld or unavailable). \\
\textcolor{blue}{The dataset does not contain all possible instances as it is impossible to capture all problems that can be written in self-contained Python code. However, the 602 instances in CodeEval dataset is diversified enough to be a good representative of a benchmark dataset to evaluate LLMs understandability of Python programming language.}
\item What data does each instance consist of? “Raw” data (e.g., unprocessed text or images) or features? In either case, please provide a description \\
\textcolor{blue}{Each instance of the dataset consists of a text formatted in JSON. The JSON text consists of a natural language description of a Python problem, the problem's solution code in Python, one of the 24 categories the problem belongs to, the complexity level of problem (beginner, intermediate, advanced), and the test cases for evaluating functional correctness of a model generated code for this problem.}
\item Is there a label or target associated with each instance? If so, please provide a description. \\
\textcolor{blue}{No.}
\item Is any information missing from individual instances? If so, please provide a description, explaining why this information is missing (e.g., because it was unavailable). This does not include intentionally removed
information, but might include, e.g., redacted text. \\
\textcolor{blue}{No.}
\item Are relationships between individual instances made explicit (e.g., users’ movie ratings, social network links)? If so, please describe how these relationships are made explicit. \\
\textcolor{blue}{The instances of the dataset are independent of each other.}
\item Are there recommended data splits (e.g., training, development/validation, testing)? If so, please provide a description of these splits, explaining the rationale behind them. \\
\textcolor{blue}{The dataset is a benchmarking dataset and the entirety of the dataset is indended to be used as a test set for evaluation of LLMs trained on code.}
\item Are there any errors, sources of noise, or redundancies in the dataset? If so, please provide a description. \\
\textcolor{blue}{N/A.}
\item Is the dataset self-contained, or does it link to or otherwise rely on external resources (e.g., websites, tweets, other datasets)? If it links to or relies on external resources, a) are there guarantees that they will exist, and remain constant, over time; b) are there official archival versions of the complete dataset (i.e., including the external resources as they existed at the time the dataset was created); c) are there any restrictions (e.g., licenses, fees) associated with any of the external resources that might apply to a future user? Please provide descriptions of all external resources and any restrictions associated with them, as well as links or other access points, as appropriate. \\
\textcolor{blue}{Yes, the dataset is self-contained.}
\item Does the dataset contain data that might be considered confidential (e.g., data that is protected by legal privilege or by doctor–
patient confidentiality, data that includes the content of individuals’ non-public communications)? If so, please provide a description \\
\textcolor{blue}{No.}
\item Does the dataset contain data that, if viewed directly, might be offensive, insulting, threatening, or might otherwise cause anxiety?
If so, please describe why. \\
\textcolor{blue}{No.}
\end{enumerate}

\subsection{Collection Process}
\begin{enumerate}
   \item How was the data associated with each instance acquired? Was the data directly observable (e.g., raw text, movie ratings), reported by subjects (e.g., survey responses), or indirectly inferred/derived from other data (e.g., part-of-speech tags, model-based guesses for age or language)? If data was reported by subjects or indirectly inferred/derived from other data, was the data validated/verified? If so, please describe how. \\
    \textcolor{blue}{Each instance of the dataset were hand-curated by the authors.}
    \item What mechanisms or procedures were used to collect the data (e.g., hardware apparatus or sensor, manual human curation, software program, software API)? How were these mechanisms or procedures validated? \\
    \textcolor{blue}{N/A.}
    \item If the dataset is a sample from a larger set, what was the sampling strategy (e.g., deterministic, probabilistic with specific sampling probabilities)? \\
    \textcolor{blue}{N/A.}
    \item Who was involved in the data collection process (e.g., students, crowdworkers, contractors) and how were they compensated (e.g., how much were crowdworkers paid)? \\
    \textcolor{blue}{Dataset development was done by the authors.}
    \item Over what timeframe was the data collected? Does this timeframe match the creation timeframe of the data associated with the instances (e.g., recent crawl of old news articles)? If not, please describe the timeframe in which the data associated with the instances was created. \\
    \textcolor{blue}{The dataset was developed between Nov 2023 - July 2025.}
    \item Were any ethical review processes conducted (e.g., by an institutional review board)? If so, please provide a description of these review processes, including the outcomes, as well as a link or other access point to any supporting documentation. \\
    \textcolor{blue}{No.}
\end{enumerate}

\subsection{Data Preprocessing}

\begin{enumerate}
    \item Was any preprocessing/cleaning/labeling of the data done (e.g., discretization or bucketing, tokenization, part-of-speech tagging, SIFT feature extraction, removal of instances, processing of missing values)? If so, please provide a description. If not, you may skip the remainder of the questions in this section. \\
    \textcolor{blue}{The labeling of the dataset into problem categories and complexity levels was performed directly by the authors. Category assignment followed pedagogical foundations (primarily Ramalho's Fluent Python) supplemented with the authors' professional expertise in Python. Complexity levels were assigned based on pedagogical principles and conceptual difficulty of required Python constructs. These assignments were then empirically validated using performance data from 15 diverse LLMs, which confirmed strong negative correlation between complexity level and model performance (overall: r = –0.324, p < 0.05; mean individual model: r = –0.829 ± 0.394) and statistically significant performance differences between levels (Cohen's d = 0.790).}
    \item Was the “raw” data saved in addition to the preprocessed/cleaned/labeled data (e.g., to support unanticipated future uses)? If so, please provide a link or other access point to the “raw” data. \\
    \textcolor{blue}{The raw data is saved in local computer and is not shared as we don't anticipate any usages of it.}
    \item Is the software used to preprocess/clean/label the instances available? If so, please provide a link or other access point. \\
    \textcolor{blue}{N/A. The intended use of CodeEval does not require pre-processing, labeling or cleaning.}
    \item Any other comments \\
    \textcolor{blue}{None.}
\end{enumerate}

\subsection{Uses}
\begin{enumerate}
    \item Has the dataset been used for any tasks already? If so, please provide a description \\
     \textcolor{blue}{The dataset has not been used for any task other than what is studied in the paper.}
     \item Is there a repository that links to any or all papers or systems that use the dataset? If so, please provide a link or other access point. \\
     \textcolor{blue}{N/A.}
     \item What (other) tasks could the dataset be used for? \\
     \textcolor{blue}{The dataset in its current form cannot be used for any other task. However, the dataset can be augmented by research communities to evaluate code-trained LLMs on several other tasks such as Code Clone Detection, Code Quality Assessment among others.}
     \item Is there anything about the composition of the dataset or the way it was collected and preprocessed/cleaned/labeled that might impact future uses? For example, is there anything that a dataset consumer might need to know to avoid uses that could result in unfair treatment of individuals or groups (e.g., stereotyping, quality of service issues) or other risks or harms (e.g., legal risks, financial harms)? If so, please provide a description. Is there anything a dataset consumer could do to mitigate these risks or harms? \\
     \textcolor{blue}{No.}
     \item Are there tasks for which the dataset should not be used? If so, please provide a description. \\
     \textcolor{blue}{No.}
     \item Any other comments? \\
     \textcolor{blue}{None.}
\end{enumerate}

\subsection{Distribution}

\begin{enumerate}
    \item Will the dataset be distributed to third parties outside of the entity (e.g., company, institution, organization) on behalf of which the dataset was created? If so, please provide a description. \\
    \textcolor{blue}{Yes, the dataset is freely and publicly available and accessible.}
    \item How will the dataset be distributed? (e.g., tarball on website, API, GitHub; does the data have a DOI and is it archived redundantly?) \\
    \textcolor{blue}{The doi of the dataset is \url{https://doi.org/10.5281/zenodo.17495202}.}
    \item When will the dataset be distributed? \\
     \textcolor{blue}{The dataset is distributed as of November 2025 in its first version 1.0.0.}
     \item Will the dataset be distributed under a copyright or other intellectual property (IP) license, and/or under applicable terms of use (ToU)? If so, please describe this license and/or ToU, and provide a link or other access point to, or otherwise reproduce, any relevant licensing terms or ToU, as well as any fees associated with these restrictions. \\
     \textcolor{blue}{The dataset is licensed under CC BY license.}
     \item Have any third parties imposed IP-based or other restrictions on the data associated with the instances? If so, please describe these restrictions, and provide a link or other access point to, or otherwise reproduce, any relevant licensing terms, as well as any fees associated with these restrictions. \\
     \textcolor{blue}{No.}
     \item Do any export controls or other regulatory restrictions apply to the dataset or to individual instances? If so, please describe these restrictions, and provide a link or other access point to, or otherwise reproduce, any supporting documentation \\
     \textcolor{blue}{No.}
     \item Any other comments? \\
     \textcolor{blue}{None.}
\end{enumerate}

\subsection{Maintenance}

\begin{enumerate}
    \item Who will be supporting/hosting/maintaining the dataset? \\
    \textcolor{blue}{The dataset is being maintained by the Computer Vision and Social Robotics Laboratory of the University of Denver.}
    \item How can the owner/curator/manager of the dataset be contacted (e.g., email address)? \\
    \textcolor{blue}{The manager of the dataset can be reached at \href{mailto:danny.brahman@du.edu}{danny.brahman@du.edu}.}
    \item Is there an erratum? If so, please provide a link or other access point. \\
    \textcolor{blue}{Currently, there is no erratum. If errors are encountered, the dataset will be updated with a new version whose links will be provided at in  the Github repository \url{https://github.com/dannybrahman/runcodeeval}.}
    \item Will the dataset be updated (e.g., to correct labeling errors, add new instances, delete instances)? If so, please describe how often, by whom, and how updates will be communicated to dataset consumers (e.g., mailing list, GitHub)? \\
    \textcolor{blue}{Same as above.}
    \item If the dataset relates to people, are there applicable limits on the retention of the data associated with the instances (e.g., were the individuals in question told that their data would be retained for a fixed period of time and then deleted)? If so, please describe these limits and explain how they will be enforced. \\
    \textcolor{blue}{N/A.}
    \item Will older versions of the dataset continue to supported/hosted/maintained? \\
    \textcolor{blue}{Yes.}
    \item If others want to extend/augment/build on/contribute to the dataset, is there a mechanism for them to do so? If so, please provide a description. Will these contributions be validated/verified? If so, please describe how. If not, why not? Is there a process for communicating/distributing these contributions to dataset consumers? If so, please provide a description. \\
    \textcolor{blue}{Extending the dataset will simply involve adding more unique problems and their corresponding test cases along with a canonical solution code.}
    \item Any other comments? \\
    \textcolor{blue}{None.}
\end{enumerate}
\end{document}